\documentclass[amsmath, amssymb, aps, pra, twocolumn,floatfix,nofootinbib]{revtex4}
\usepackage[utf8]{inputenc}
\usepackage{multirow}
\usepackage[urlcolor=blue, colorlinks=true,citecolor=blue]{hyperref}
\usepackage{enumitem}

\usepackage{amsmath}
\usepackage{amsmath, amssymb}
\usepackage{textcomp, gensymb}
\usepackage{tabularx}

\usepackage{wrapfig}
\usepackage{graphicx}
\usepackage{float}
\usepackage{subfigure}
\usepackage{amssymb}
\usepackage{bbold}
\usepackage{color}
\usepackage{ulem}

\newcommand{\ket}[1]{\left|#1\right\rangle}
\newcommand{\bra}[1]{\left\langle #1\right|}

\preprint{APS/123-QED}
\begin{document}
\title{Quantum machine learning for image classification}

\author{Arsenii Senokosov}
\author{Alexandr Sedykh}
\author{Asel Sagingalieva}
\author{Basil Kyriacou}
\author{Alexey Melnikov}
\thanks{Corresponding author, e-mail: alexey@melnikov.info
\begin{center}
\fbox{
\begin{minipage}{0.45\textwidth}
Please check the published version, which includes all the latest additions and corrections: Mach. Learn.: Sci. Technol. 5:015040, 2024, DOI: \href{https://doi.org/10.1088/2632-2153/ad2aef}{10.1088/2632-2153/ad2aef}
\end{minipage}
}
\end{center}
}

\affiliation{Terra Quantum AG, Kornhausstrasse 25, 9000 St.~Gallen, Switzerland}


\begin{abstract}
Image classification, a pivotal task in multiple industries, faces computational challenges due to the burgeoning volume of visual data. This research addresses these challenges by introducing two quantum machine learning models that leverage the principles of quantum mechanics for effective computations. Our first model, a hybrid quantum neural network with parallel quantum circuits, enables the execution of computations even in the noisy intermediate-scale quantum era, where circuits with a large number of qubits are currently infeasible. This model demonstrated a record-breaking classification accuracy of $99.21\%$ on the full MNIST dataset, surpassing the performance of known quantum–classical models, while having eight times fewer parameters than its classical counterpart. Also, the results of testing this hybrid model on a Medical MNIST (classification accuracy over $99\%$), and on CIFAR-10 (classification accuracy over $82\%$), can serve as evidence of the generalizability of the model and highlights the efficiency of quantum layers in distinguishing common features of input data. Our second model introduces a hybrid quantum neural network with a Quanvolutional layer, reducing image resolution via a convolution process. The model matches the performance of its classical counterpart, having four times fewer trainable parameters, and outperforms a classical model with equal weight parameters. These models represent advancements in quantum machine learning research and illuminate the path towards more accurate image classification systems.

\end{abstract}

\maketitle

\section*{Introduction}

Image classification is a critical task in the modern world due to its wide range of practical applications in various fields~\cite{10.48550/arxiv.1506.01497}. For instance, in medical imaging, image classification algorithms have been shown to significantly improve the accuracy and speed of diagnoses of many diseases~\cite{litjens2017survey, zhou2020review}. In the field of autonomous vehicles, image classification plays a crucial role in object detection, tracking, and classification, which is necessary for safe and efficient navigation.

Deep learning approaches~\cite{lecun2015deep} like deep convolutional neural networks (CNNs) have emerged as powerful tools for image classification and recognition tasks~\cite{segnet, densenet}, achieving state-of-the-art performance on various benchmark datasets~\cite{krizhevsky2012imagenet, redmon2017yolo9000}. However, as the amount of visual data grows, modern neural networks face significant computational challenges.

 Quantum technologies, on the other hand, offer the potential to overcome this computational limitation by harnessing the power of quantum mechanics to perform computations in parallel~\cite{ladd2010quantum}. Quantum machine learning (QML) is a rapidly evolving field that combines the principles of quantum mechanics, and classical machine learning~\cite{Schuld_2014, Biamonte2017, Dunjko.Briegel.2018, qml_review_2023}. This field has the potential to revolutionize various areas of computing, including image classification~\cite{cong2019quantum, hur2022quantum, herrmann2022realizing}. It has attracted significant attention due to its potential to solve computational problems that classical computers are unable to solve efficiently~\cite{ladd2010quantum}. This potential arises from the unique features of quantum computing, such as superposition and entanglement, which can provide an exponential speedup for specific machine learning tasks~\cite{Schuld2018}. QML algorithms produce probabilistic results, which align well with classification problems~\cite{fphy_2022_1069985}. They also operate in an exponentially larger search space, which has the potential to enhance their performance~\cite{Lloyd2013, 365700, Lloyd1996UniversalQS}. However, it's important to note that the realization of these advantages in practical applications remains an active area of research and investigation. However, the real-world implementation of quantum algorithms faces significant challenges, such as the need for error correction and the high sensitivity of quantum systems to external disturbances~\cite{Aaronson2015}. Despite these challenges, QML has shown promising results in several applications~\cite{Wiebe2015}. In the context of image classification, QML algorithms can process large datasets of images more efficiently than classical algorithms, leading to faster and more accurate classification~\cite{Havlicek2019}. Recent studies have also explored hybrid quantum-classical convolutional neural networks and demonstrated the classification~\cite{liu2021hybrid, li2022image, houssein2022hybrid, fphy_2022_1069985, 10.1016/j.optcom.2023.129287, sagingalieva2023hyperparameter, 10.1063/5.0138021, 10.1093/jcde/qwac003} and generation~\cite{zhou2023hybrid,tsang2022hybrid,huang2021experimental,rudolph2022generation} of images.

A promising area of research within QML for image classification is the hybrid quantum neural network (HQNN)~\cite{Schuld2018,boston-housing, 10.48550/arxiv.2109.02862, 10.1109/idaacs53288.2021.9661011}. HQNNs combine classical deep learning architectures with QML algorithms~\cite{marshall2022high,kordzanganeh2023exponentially,perez2022reduce,kordzanganeh2023parallel,jerbi2023quantum}, namely Parameterized Quantum Circuits (PQCs), creating a hybrid system that leverages the strengths of both classical and quantum computing. This approach allows for the processing of large datasets with greater efficiency than classical deep learning architectures alone~\cite{li2020quantum, Mitarai2020}. HQNNs have shown promise in a variety of industrial tasks, e.g., in the healthcare~\cite{10.1063/5.0138021,houssein2022hybrid,domingo2023hybrid,pharma,jain2022hybrid}, chemical~\cite{sedykh2023quantum, kurkin2023forecasting}, financial~\cite{10.1016/j.eswa.2022.116583}, and aerospace industries~\cite{rainjonneau2023quantum, haboury2023supervised}. Further research is needed to explore the full potential of HQNNs in image classification and to develop more robust and scalable algorithms.

In this article, we propose two approaches to leverage quantum computing in the field of image recognition. The first approach involves applying parallel PQCs after classical deep convolutional layers, while the second approach involves using an HQNN with a quanvolutional layer. We evaluate the performance of these hybrid models on the MNIST dataset of hand-written digits, which is described in Section~\ref{MNIST_dataset}, and demonstrate their ability to classify images.

Previous studies have primarily focused on either purely quantum solutions~\cite{Farhi.Neven.2018mcl, Baek.Kim.2022, chen2023quantum} for image recognition or various hybrid models~\cite{10.48550/arxiv.2109.02862}. Yet, the specific potential of seamlessly integrating quantum circuits with classical neural networks remains an under-explored area. In this work, we venture into this niche and present an innovative architecture for the HQNN, that is specifically designed to operate even in the NISQ era, setting a benchmark by achieving record-breaking classification accuracy with significantly fewer parameters than its classical counterpart. Importantly, we achieved these results without resorting to pretrained models or transfer learning techniques, underscoring the inherent strength of our model's design and training process. Our successful test results across multiple datasets further support the claim of the model's broad applicability and generalizability.

The first model (described in Section~\ref{first_approach}) combines classical convolutional layers with parallel quantum layers (HQNN-Parallel). The quantum part is analogous to a classical fully connected layer. We compare the hybrid model with its most closely corresponding classical counterpart (in terms of the architecture and the number of layers) and observe that the hybrid model outperforms the classical model in accuracy (achieving $99.21\%$ accuracy) despite having eight times fewer parameters. Moreover, we tested this model on two more datasets, on Medical MNIST, which is described in Section~\ref{Medical_MNIST}, and on CIFAR-10, which is described in Section~\ref{CIFAR-10}, to ensure its generalizability. 

In the second model (described in Section~\ref{second_approach}), we introduce HQNN with a quanvolutional layer (HQNN-Quanv), which is a kernel that applies a convolution to the input image and reduces its resolution. The HQNN-Quanv achieves similar accuracy to the classical model ($67\%$ accuracy) despite having four times fewer trainable parameters in the first layer compared to the classical counterpart. Additionally, the hybrid model outperforms the classical model with the same number of weights.

We note that having fewer trainable parameters does not necessarily lead to a more efficient execution of quantum machine learning models compared to classical ones. The reason for this is the much more expensive training and operation costs required for quantum models relative to classical models. However, with future advancements in quantum hardware technologies, the gap in training and operation costs between quantum and classical models might become narrower. This highlights the potential of quantum computing and QML in advancing the field of image recognition. Our results contribute to the ongoing research in this area and demonstrate the exciting possibilities for the future of QML in other fields.

\section*{Results}\label{results}

\subsection{Datasets}

\subsubsection{$\mathrm{MNIST}$}\label{MNIST_dataset}

\begin{figure}[ht]
    \centering
    \includegraphics[width=\linewidth]{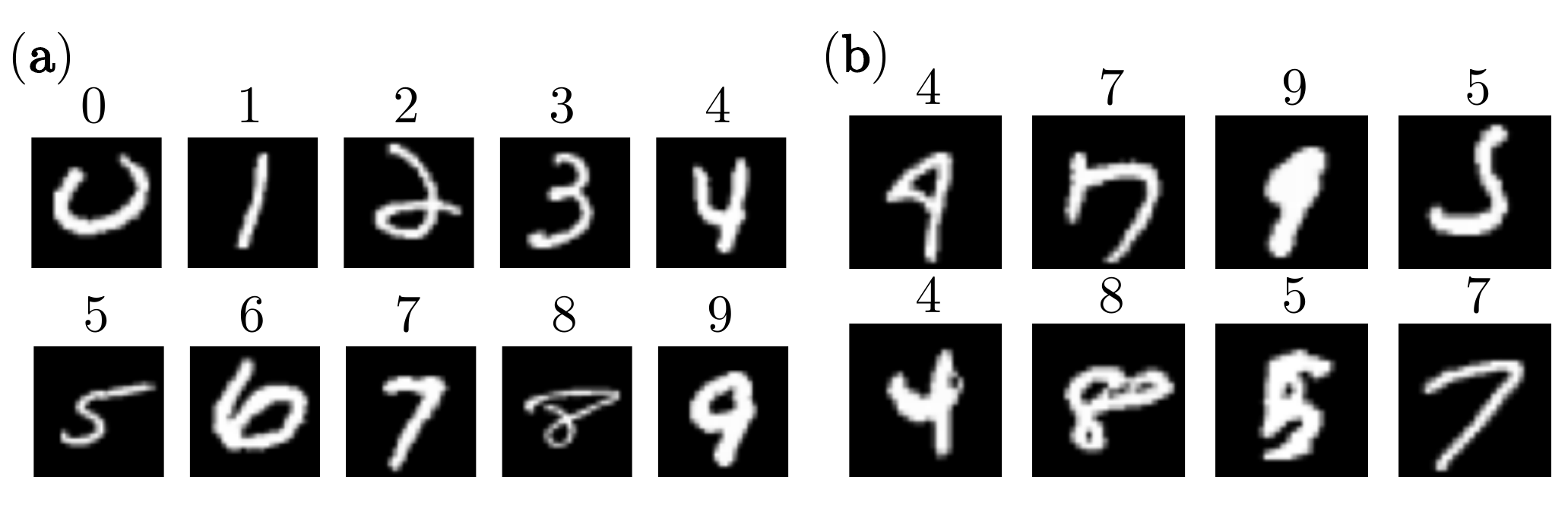}
    \caption{a) Examples of images from the MNIST dataset. b)~Examples of ambiguous images from the MNIST dataset.}
    \label{fig:MNIST}
\end{figure}

This section describes the Modified National Institute of Standards and Technology (MNIST)~\cite{MNIST} dataset. The MNIST database consists of a large collection of grey-scale handwritten numbers, ranging from $0$ to $9$. Sample images from the dataset are presented in Fig.~\ref{fig:MNIST}(a). Each image has a resolution of $28\times28$ pixels, and the main objective is to classify each image by assigning a class label using a neural network. In other words, the task is to recognize which digit is present in the image. This dataset is widely used for making the first steps in the sphere of machine learning. Nevertheless, it is worth studying as it helps to test the performance of various neural network models~\cite{Cire_an_2010, 10.1109/cvpr.2012.6248110}, especially models with PQCs~\cite{10.48550/arxiv.2109.02862, Farhi.Neven.2018mcl, 10.48550/arxiv.2303.05860}. The MNIST dataset used in this study comprises a total of $70000$ images, with $60000$ images reserved for training and $10000$ images for testing. However, in certain cases, it may be advantageous to reduce the number of images in order to expedite the training process and gain immediate insights into the model's performance.

Despite being a widely used dataset, the MNIST database contains a few images that are broken or ambiguous, posing a challenge even for human evaluators. Fig.~\ref{fig:MNIST}(b) provides examples of such images. However, our hybrid model can accurately determine the number in such images with over $99\%$ accuracy.

\subsubsection{Medical $\mathrm{MNIST}$}\label{Medical_MNIST}

\begin{figure}[ht]
    \centering
    \includegraphics[width=1\linewidth]{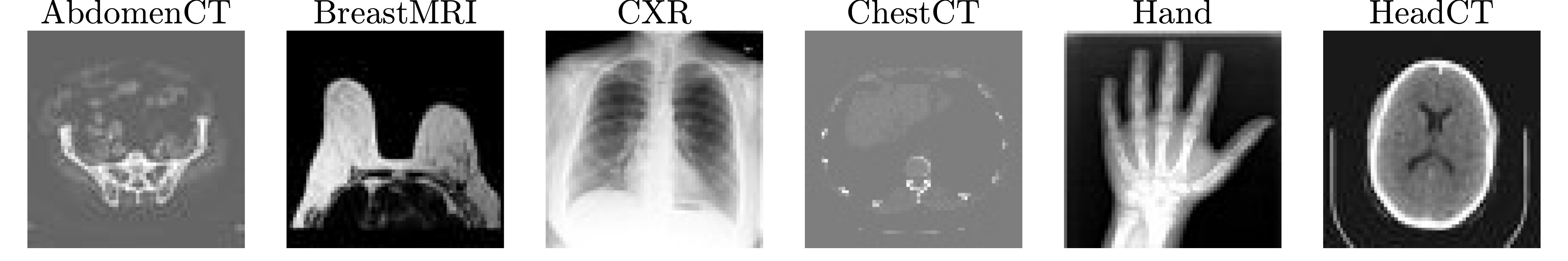}
    \caption{Examples of images from the Medical MNIST dataset.}
    \label{fig:MedNIST}
\end{figure}

The integration of QML techniques, such as HQNNs, holds immense promise in the field of medical image classification. Quantum computing's inherent capacity to handle complex and high-dimensional data, coupled with the power of neural networks, offers a unique advantage for solving intricate medical image analysis tasks. HQNNs leverage quantum computing's ability to efficiently perform certain mathematical operations required for image feature extraction and classification, potentially leading to breakthroughs in medical diagnosis and treatment~\cite{landman2022quantum}.

To test our models in this sphere, we utilized the Medical MNIST dataset~\cite{MedicalMNIST}, which comprises a total of $58954$ medical images distributed across six distinct categories: Abdomen Computer Tomography (AbdomenCT), Breast Magnetic Resonance Imaging (BreastMRI), Chest X-Ray (CXR), Chest Computer Tomography (ChestCT), Hand X-Ray (Hand), and Head Computer Tomography (HeadCT). Sample images from this dataset can be observed in Figure~\ref{fig:MedNIST}. Each of these images possesses a resolution of $64 \times 64$ pixels, with our primary goal being the classification of each image through the utilization of a neural network.

It is noteworthy that all images within this dataset employ 3 channels, adding an additional layer of complexity compared to the original MNIST dataset. Furthermore, we conducted various data preprocessing techniques, including random rotations of up to $10$ degrees, random horizontal flips, and resizing to dimensions as large as $244 \times 244$ pixels. These steps were implemented to stabilize the training process and enhance the performance of our models.

Similar to the MNIST classification task, we divided the entire dataset into two subsets: a training set consisting of $47163$ samples and a testing set containing $11791$ samples.

\subsubsection{$\mathrm{CIFAR}$-$10$}\label{CIFAR-10}

\begin{figure}[ht]
    \centering
    \includegraphics[width=\linewidth]{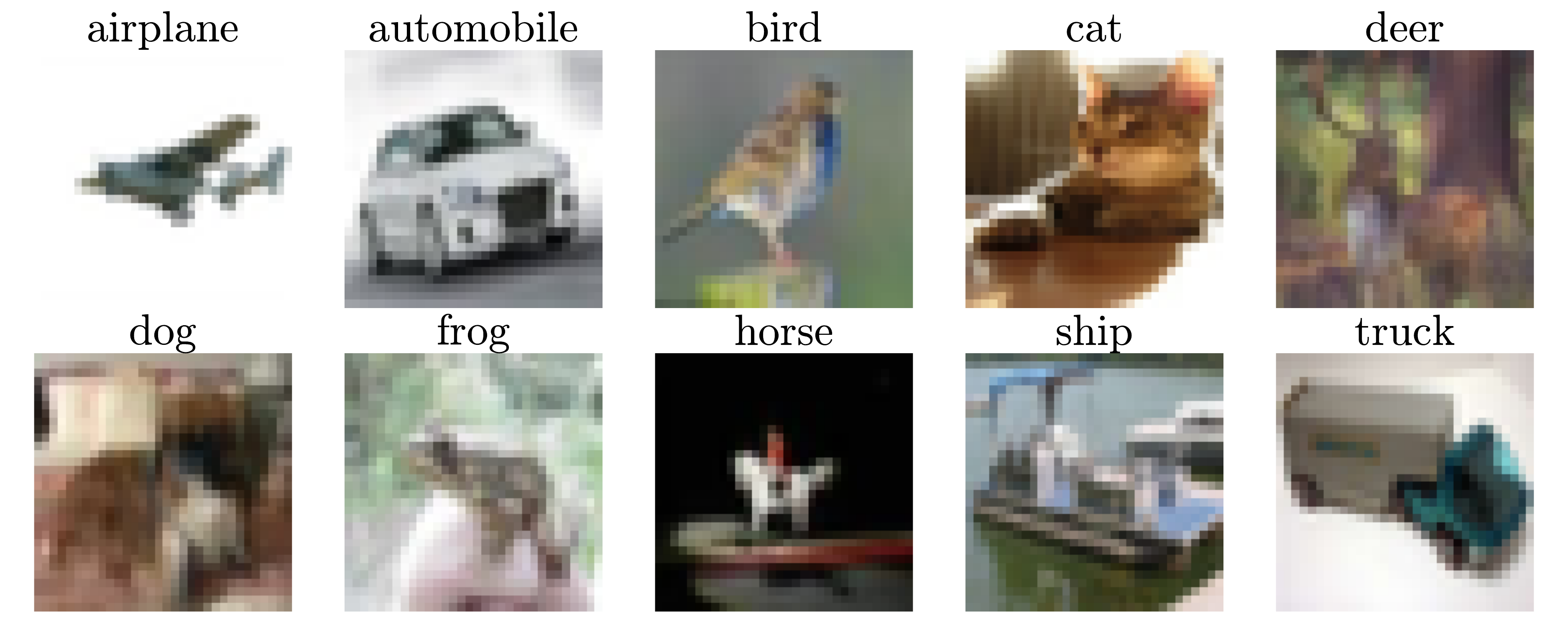}
    \caption{Examples of images from the CIFAR-10 dataset.}
    \label{fig:cifar}
\end{figure}

The CIFAR-10 dataset is a pivotal resource in the domain of computer vision and image classification. Developed as part of the Canadian Institute for Advanced Research (CIFAR) program, this dataset serves as a fundamental benchmark for assessing the performance of machine learning algorithms in the context of image classification tasks~\cite{Krizhevsky09}. CIFAR-10 is comprised of a total of $60000$ images, which are divided into training and testing sets with $50000$ and $10000$ samples in each respectively. Each image in the CIFAR-10 dataset is of size $32\time 32$ pixels. These images are color images, incorporating three color channels: red, green, and blue. Consequently, every image is represented as a $32\times 32\times 3$ tensor. One of the distinguishing features of CIFAR-10 is its categorization into ten distinct classes, each representing a different object category. These classes are as follows: airplane, automobile, bird, cat, deer, dog, frog, horse, ship, truck. This diverse set of classes ensures that the dataset is suitable for a wide range of image classification challenges, spanning various domains and object types. As it has more classes than Medical MNIST and input images have bigger resolution than in MNIST, classifying images from CIFAR-10 seems to be complex task. Sample images from this dataset are shown in Fig.~\ref{fig:cifar}.

\subsection{Hybrid Quantum Neural Network with parallel quantum dense layers, HQNN-Parallel}\label{first_approach}

\begin{figure*}[ht]
    \centering
    \includegraphics[width=1\linewidth]{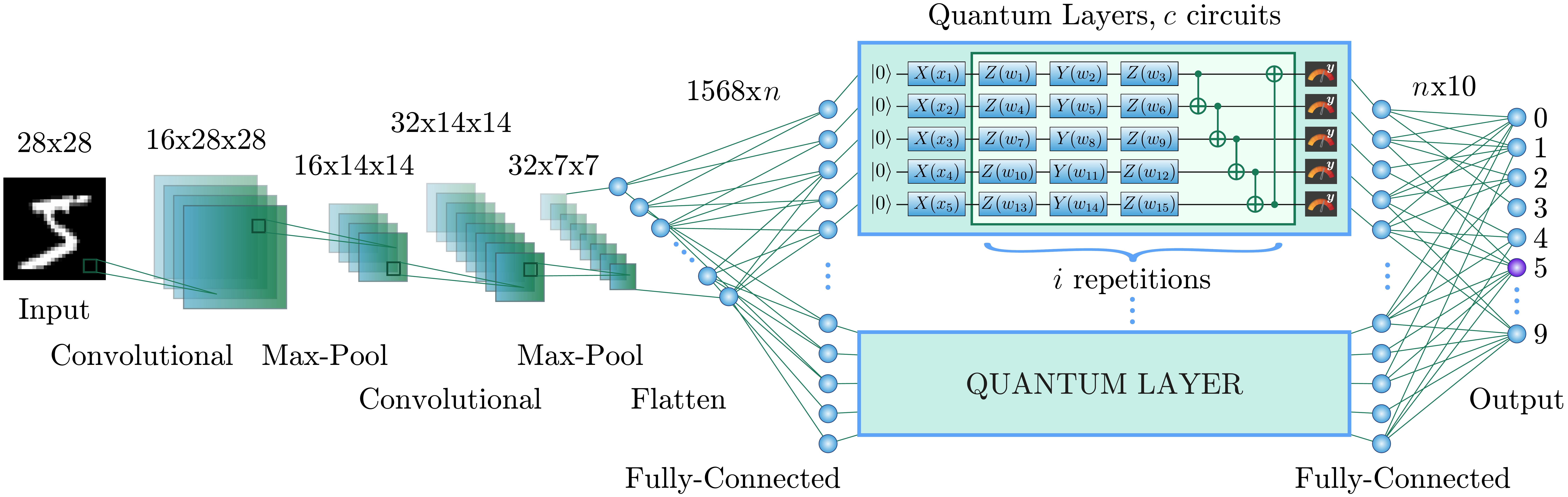}
    \caption{Architecture of the proposed HQNN-Parallel. The input data samples are transformed by a series of convolutional layers, that extract relevant features and reduce the dimensionality of the input. The output channels of the convolutional layers are then flattened into a single vector before being fed into the dense part of the HQNN-Parallel. The hybrid dense part contains a combination of classical and quantum layers. The quantum layers are implemented using parallel PQCs, which allow for simultaneous execution, reducing the total computation time. Quantum layers are depicted in the figure as blue rectangles, the top rectangle is a detailed version of subsequent quantum layers in the amount of $c$ circuits. The output of the last classical fully connected layer is a predicted digit in the range of 0 to 9.}
    \label{fig:qfc}
\end{figure*}

This section describes our first proposed model, the Hybrid Quantum Neural Network with parallel quantum dense layers, each of which is a PQC. Section~\ref{quantum_layer} presents the results of our comparison between the hybrid model and its classical counterpart, CNN~\ref{trainingresults_1}.~The HQNN-Parallel consists of two main components: a classical convolutional block~\ref{conv} and a combination of classical fully connected and parallel quantum layers \ref{dense}.~The main purpose of the classical convolutional block is to reduce the dimensionality of the input data and prepare it for further processing. The classical fully connected as well as parallel quantum layers constitute the core of the HQNN-Parallel, and are responsible for prediction tasks of the model. Further details on the architecture and implementation of the HQNN-Parallel are presented in subsequent sections.

\subsubsection{Classical Convolutional Layers}\label{conv}

Fig.~\ref{fig:qfc} depicts the general structure of the classical convolutional part of the proposed HQNN-Parallel. The convolutional part of the network is comprised of two main blocks, followed by fully-connected layers. In this study, we utilized Rectified Linear Unit (ReLU) as the activation function~\cite{Agarap2018DeepLU}. Batch Normalization~\cite{https://doi.org/10.48550/arxiv.1502.03167} is employed in the network as it stabilizes the training process and improves the accuracy of the model.

The first block of the convolutional part of the HQNN-Parallel comprises a convolutional layer with one input channel and $16$ output channels, utilizing a square kernel of size $5\times5$. The layer operates with a stride of one pixel and applies a two-pixel padding to the input data. We opted for the $5\times5$ kernel size primarily to maintain the input image's spatial dimensions. By implementing a two-pixel padding, the $5\times5$ kernel facilitates the creation of $16$ channels without altering the original pixel size of $28\times28$. Additionally, the larger kernel sizes, such as $5\times5$, can capture more complex and diverse features from the input image compared to smaller kernels~\cite{tan2019mixconv, tan2019mnasnet}. This aligns with our objective of preserving both spatial and feature details of the image during convolution. Batch Normalization is applied to the output of the convolutional layer, followed by an activation function (ReLU) and MaxPooling~\cite{lecun1998gradient} with a kernel size of two pixels. The resulting feature map has dimensions of $16\times14\times14$ pixels.

The second block contains a convolutional layer with $16$ input channels and $32$ output channels, utilizing the same kernel size and padding as the previous layer. The MaxPooling parameters remain unchanged, resulting in a feature map with dimensions of $32\times7\times7$ pixels, which will become an input for the fully connected part of the network.

\subsubsection{Hybrid Dense Layers}\label{dense}

Following the convolutional block, the HQNN-
Parallel, continues with a hybrid dense part, as shown in Fig. \ref{fig:qfc}. The $32\times7\times7$ feature map produced by the convolutional part serves as input for the first dense layer, which transforms the feature map from $1568$ to $n$ features. The value of $n$ is determined by the chosen quantum part and represents the total number of encoding parameters in the quantum layers.

Each quantum layer is designed to maintain the number of input and output features, and the output of the quantum layer is fed into the second classical fully connected layer. This layer performs the final transformation and maps the $n$ input features to $10$ output features, corresponding to the number of classes into which the images can be classified. After each classical dense layer, Batch Normalization and ReLU activation are applied.

It is worth noting that the structure of the HQNN-
Parallel, including the number of layers and the number of features, can be adjusted to optimize the performance on a specific task.

\subsubsection{Structure of Quantum Layer}\label{quantum_layer}

\begin{figure*}[ht]
    \centering
    \includegraphics[width=1\linewidth]{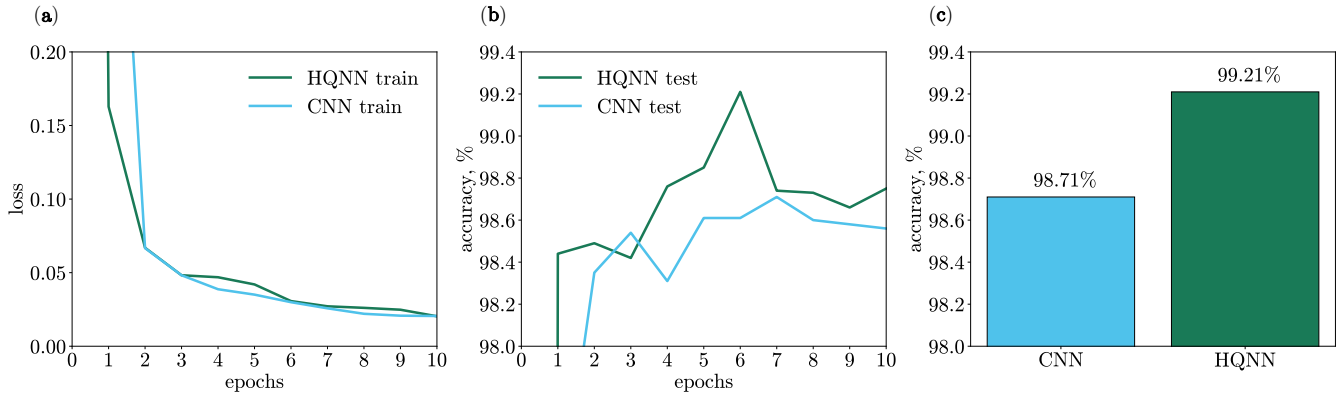}
    \caption{(a-b) Train and test results for the HQNN-Parallel and the CNN. The HQNN has a 99.21\% accuracy on the test data and outperforms the CNN which has a 98.71\% accuracy. The classical model has 8 times more variational parameters than the hybrid one. (c) Test accuracies of the HQNN-Parallel and its classical counterpart, the CNN.}
    \label{fig:classical}
\end{figure*}

\begin{figure*}
    \centering
    \includegraphics[width=1\linewidth]{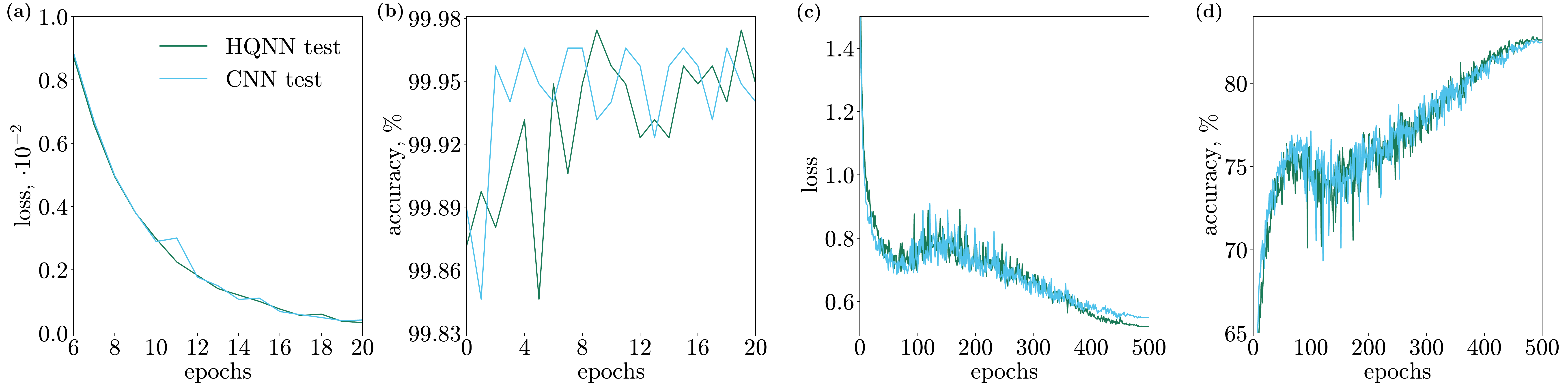}
    \caption{(a-b) Test results for the HQNN-Parallel and the CNN on Medical MNIST dataset. The HQNN has a $99.97\%$ accuracy on the test data and slightly outperforms the CNN which has a $99.96\%$ accuracy. (c-d) Test results for the HQNN-Parallel and the CNN on CIFAR-10 dataset. The HQNN has a $82.78\%$ accuracy on the test data and outperforms the CNN which has a $82.64\%$ accuracy.}
    \label{fig:Pharma-MNIST}
\end{figure*}

The quantum component of the proposed HQNN-Parallel, depicted in Fig.~\ref{fig:qfc}, consists of $c$ parallel quantum layers, each of which is a PQC composed of three parts: embedding, variational gates, and measurement. The input data to the quantum layers are $n$ features from the previous classical fully connected layer, divided into $c$ parts, with each part being a vector of $q$ values, $x =(\phi_1, \phi_2, ..., \phi_q) \in \mathbb{R}^q$. To encode these classical features into quantum Hilbert space, we use the ``angle embedding'' method, which rotates each qubit in the ground state around the X-axis on the Bloch sphere~\cite{PhysRev.70.460} by an angle proportional to the corresponding value in the input vector: $\ket{\psi} = R^\mathrm{emb}_x(x)\ket{\psi_0}$, where $\ket{\psi_0} = \ket{0}^{\otimes q}$. This operation encodes the input vector into quantum space, and the resulting quantum state represents the input data from the previous classical layer. It is important to note that $n$ is divisible by $q$ since the input data vector is divided into $c = n/q$ parts, with each part serving as input to a PQC.

The encoding part for each PQC is followed by a variational part, which consists of two parts: rotations with trainable parameters and subsequent CNOT operations~\cite{Barenco_1995}. The rotations serve as quantum gates that transform the encoded input data according to the variational parameters, while the CNOT operations entangle the qubits in the PQC. The depth of the variational part, denoted as $i$, is a hyperparameter that determines the number of iterations of the rotations and CNOT operations in the PQC. It is important to note that the variational parameters for each PQC are different in each of the $i$ repetitions and for each of $c$ quantum circuits. Thus, the total number of weights in the quantum part of the HQNN-Parallel is calculated as $q\cdot 3i\cdot c$.

After performing these operations, measurement in the Pauli basis matrices is performed, resulting in 
\begin{equation}
v^{(j)} = \bra{0} {R^\mathrm{emb}_x(\phi_j)}^\dagger {U(\theta)}^\dagger Y_j U(\theta) R^\mathrm{emb}_x(\phi_j)\ket{0},    
\end{equation} where $Y_j$ is the Pauli-Y matrix for the $j^\text{th}$ qubit, $R^\mathrm{emb}_x(\phi_j)$ and  $U(\theta)$  are operations, performed by the embedding and trainable parts of the PQC, respectively, and $\theta$ is a vector of trainable parameters. After this operation, we have the vector $v\in\mathbb{R}^q$. The outputs of all the PQCs would be concatenated to form a new vector $\hat{v}\in\mathbb{R}^n$ that is the input data for a subsequent classical fully-connected layer. This layer, being the final layer in the classification pipeline, produces an output in the form of a probability distribution over the set of classes. In our case, each input image is associated with one of the ten possible digits from $0$ to $9$, and the output of each neuron represents the probability that the image belongs to that class. The neuron with the highest output probability is selected as the predicted class for the image.

More detailed theoretical analysis, involving ZX-Calculus reduction and Fourier expressivity of the HQNN-Parallel was conducted in Appendix \ref{sec:analysis:theory}.

\subsubsection{Training and results}\label{trainingresults_1}

\begin{table}[ht!]
\centering

\begin{tabular}{|c|c|c|c|c|c|}
\hline
Dataset                                                                  & Model & train loss            & test loss             & \begin{tabular}[c]{@{}c@{}}test\\ acc\end{tabular} & \begin{tabular}[c]{@{}c@{}}param\\ num\end{tabular} \\ \hline
\multirow{2}{*}{MNIST}                                                   & CNN   & 0.0205                & 0.0449                & 98.71                                              & 372234                                              \\ \cline{2-6} 
                                                                         & HQNN  & 0.0204                & 0.0274                & 99.21                                              & 45194                                               \\ \hline
\multirow{2}{*}{\begin{tabular}[c]{@{}c@{}}Medical\\ MNIST\end{tabular}} & CNN   & $0.456 \cdot 10^{-2}$ & $0.396 \cdot 10^{-2}$ & 82.64                                              & 247642                                              \\ \cline{2-6} 
                                                                         & HQNN  & $0.429 \cdot 10^{-2}$ & $0.332 \cdot 10^{-2}$ & 82.78                                              & 247462                                              \\ \hline
\multirow{2}{*}{CIFAR-10}                                                & CNN   & 0.3659                & 0.5484                & 82.78                                              & 81698                                               \\ \cline{2-6} 
                                                                         & HQNN  & 0.3851                & 0.5208                & 82.64                                              & 81578                                               \\ \hline
\end{tabular}

\caption{Summary of the results for the HQNN-Parallel and its classical analog, CNN.}
\label{tab:summaryi}
\end{table}

\begin{figure*}[ht!]
    \centering
    \includegraphics[width=0.9\linewidth]{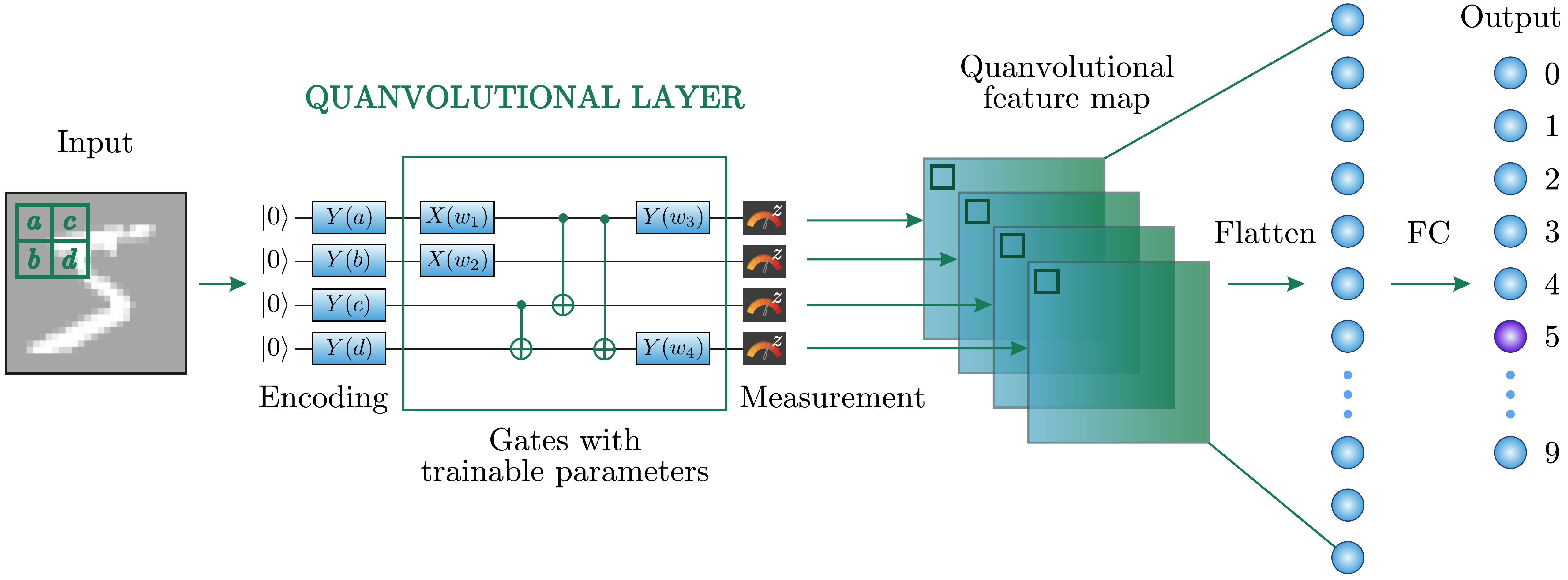}
    \caption{Architecture of the HQNN-Quanv. The Quanvolutional layer maps the input image into $4$ Quanvolutional feature maps. These feature maps are then concatenated and flattened to go into the fully-connected classical layer, which gives us $10$ probabilities for each class.}
    \label{fig:quanv}
\end{figure*}

As described above, the HQNN-Parallel was trained on MNIST dataset~\ref{MNIST_dataset}. No preprocessing is applied, so the entire collection is used for training ($60000$ images are in the training set and $10000$ are in the test set). In the context of training the proposed HQNN-Parallel, the ultimate objective is to minimize the loss function during the optimization process. The cross-entropy function is employed as the loss function, given by:
\begin{equation}
l = - \sum_{c=1}^{k}y_c\log{p_c},     
\end{equation}
where $p_c$ is the prediction probability, $y_c$ is either 0 or 1, determining respectively if the image belongs to the prediction class, and $k$ is the number of classes.

The parameters of the classical layers are optimized using the backpropagation algorithm~\cite{Rumelhart1986LearningIR}, which is automatically implemented in the PyTorch library~\cite{NEURIPS2019_9015}. The backpropagation algorithm is used to calculate the gradients of the loss function with respect to the parameters of the network, allowing for their optimization via gradient descent. However, the use of quantum layers in this task is more complex than classical methods for computing gradients. To overcome this challenge, we employ the PennyLane framework~\cite{Bergholm.Killoran.2018}, which provides access to a variety of optimization techniques. We utilize the parameter shift rule~\cite{Wierichs_2022}, which is compatible with physical implementations of quantum computing~\cite{benchmarkingQuantum}. This method involves evaluating the gradient of a quantum circuit by shifting the parameters in the circuit and computing the corresponding change in the circuit's output. The resulting gradient can then be used to update the circuit's parameters and iteratively minimize the loss function. By using the parameter shift rule, we are able to efficiently optimize the variational parameters in the quantum layers of the HQNN, enabling the network to learn complex patterns in the input data and achieve accurate results.

In the process of solving the problem, we tried various architectures of quantum layers. The most successful architecture for the HQNN-Parallel used a quantum layer with $5$ qubits and $3$ repetitions of the strongly entangling layers. The number of quantum layers is equal to $4$.

The HQNN-Parallel managed to achieve a $99.21\%$ accuracy on MNIST dataset. In order to compare the performance of the HQNN with a classical CNN, the convolutional part of the HQNN was held constant, while the quantum part was replaced with a classical dense layer containing $n$ neurons. This modified CNN was then trained on the same MNIST dataset. A comparison of the training outcomes is depicted in Fig.~\ref{fig:classical}(a-b).

The trainable parameters, as well as the primary training and testing results, for both the HQNN-Parallel and the CNN, are summarized in Table~\ref{tab:summaryi} and illustrated in Fig.~\ref{fig:classical}(c). From these results, it is evident that the most successful implementation of the HQNN-Parallel surpasses the performance of a CNN that possesses approximately eight times more parameters.

HQNN-Parallel with its classical analog were also tested on Medical MNIST dataset. The HQNN-Parallel managed to achieve a $99.97\%$ accuracy. The classical CNN showed less accurate results with $99.96\%$ accuracy on test data. It is worth noting that HQNN model had 247462 trainable parameters and classical CNN had 247642. A comparison of the training outcomes is depicted in Fig.~\ref{fig:Pharma-MNIST}(a-b).

To confirm the generalizability of hybrid architecture we tested hybrid and classical models on CIFAR-10 dataset. The HQNN-Parallel managed to achieve an $82.78\%$ accuracy. The classical CNN showed less accurate results with $82.64\%$ accuracy on test data. It is worth noting that HQNN model had $81578$ trainable parameters and classical CNN had $81698$. A comparison of the training outcomes is depicted in Fig.~\ref{fig:Pharma-MNIST}(c-d).

In this section, we provide a comprehensive overview of the model's architecture employed during training on the MNIST dataset. While the foundational structure remained consistent when testing on both the Medical MNIST and CIFAR-10 datasets, there were variations. Specifically, input dimensions varied based on image sizes. Additionally, the number of convolutions differed: two for Medical MNIST and three for CIFAR-10. Furthermore, the neuron count in the output layer was adjusted in line with the respective number of classification classes. The quantity of qubits and quantum layers remained unchanged across all experiments.

\subsection{Hybrid Quantum Neural Network with quanvolutional layer, HQNN-Quanv}\label{second_approach}

In this section, we give a detailed description of our second hybrid quantum approach for solving the problem of recognizing the numbers from the MNIST dataset, based on the combination of a quanvolutional layer and classical fully connected layers. The scheme of this network is presented in Fig.~\ref{fig:quanv}. We also compare our hybrid model with its classical analog CNN and investigate the relationship between quanvolutional and convolutional layers, as well as their dependence on the number of output channels.

\begin{figure*}[ht!]
    \centering
    \includegraphics[width=1\linewidth]{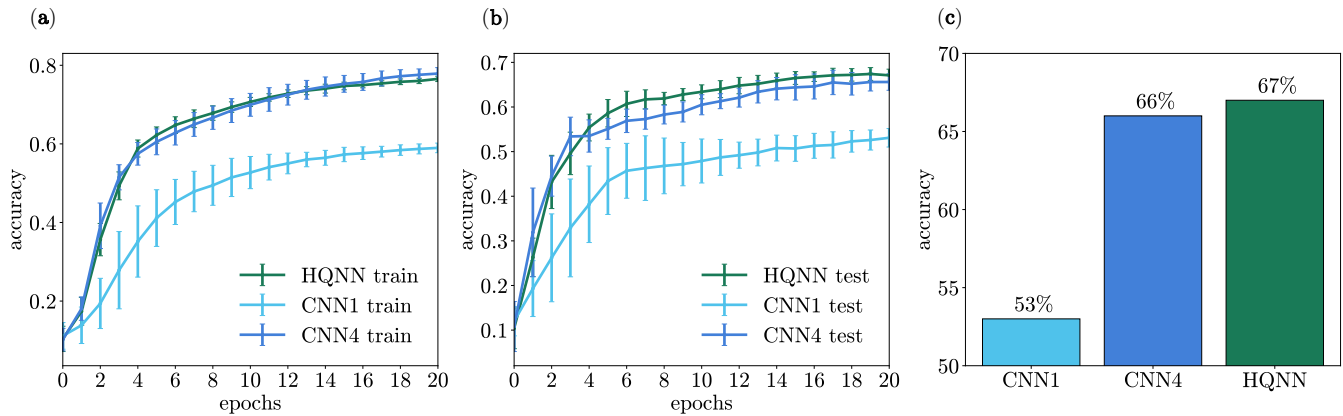}
    \caption{(a-b) Train and test accuracies for the CNN and HQNN-Quanv models with stride set to 4. The models differ only in the kernel and the number of output channels.
    1. HQNN: Quanvolutional kernel with 1 input channel, and 4 output channels;
    2. CNN1: Convolutional kernel with 1 input channel, and 1 output channel;
    3. CNN4: Convolutional kernel with 1 input channel, and 4 output channels.
    The HQNN-Quanv achieved an accuracy of $67 \pm 1\%$ on the test data, outperforming the CNN1 with an accuracy of $53 \pm 2\%$ and the CNN4 with an accuracy of $66 \pm 2\%$. Notably, the CNN1 has the same number of weights in the kernel as the hybrid model, while the CNN4 has four times more weights than the hybrid model.
    (c) Test accuracies for HQNN-Quanv ($67\%$), CNN1 ($53\%$) and CNN4 ($66\%$). The HQNN outperforms the CNN1, which has the same number of variational parameters. The HQNN's accuracy score is equivalent to CNN4's, which has four times many weights in its kernel.}
    \label{fig:quanv_accuracy}
\end{figure*}

\subsubsection{Quanvolutional layer}

The general architecture of a quanvolutional layer~\cite{Quanvolutional_networks} is shown in Fig.~\ref{fig:quanv}. Similar to classical convolutional layers, the quanvolutional layer comprises a kernel of size $n \times n$ pixels that convolve the input image, producing a lower-resolution output image. However, the quanvolutional layer is unique in the sense that its kernel is implemented using a quantum circuit consisting of $n_q$ qubits. The circuit can be decomposed into three distinct parts: classical-to-quantum data encoding, variational gates, and quantum measurement. These parts work together to determine the kernel's action on the input image.

There are plenty of encoding (embedding) methods to transfer classical data into quantum states. In this section, as in the previous one, we use the ``angle embedding'' technique. It is achieved by rotating the qubits from their initial $\ket{0}$ value with the $R_y(\varphi)$ unitaries, where $\varphi$ is determined by the value of the corresponding pixel. After the classical data is encoded, the quantum states undergo unitary transformations, defined by the variational part.

The variational part in the quanvolutional layer usually consists of arbitrary single-qubit rotations and CNOT gates, arranged in a particular way determined by the researcher. The unitaries in the PQC are parameterized by a set of variational parameters, which are learned via training the neural network. The ultimate goal of the model training is to find a measurement basis (by tweaking variational gate parameters) that tells us the most information about a fragment of a picture confined by the quantum kernel.

Finally, for each wire, the expectation value of an arbitrary operator is calculated to obtain the classical output. As it is a real number, it represents the kernel's output pixel, while each wire yields a different image channel. For instance, a quanvolutional kernel of size $2 \times 2$ has a $4$-qubit circuit, which transforms one input image into four images of reduced size.

\subsubsection{Structure of HQNN-Quanv}

This subsection details the architecture of the HQNN-Quanv, which is shown in Fig.~\ref{fig:quanv}. At first, a simple angle embedding of the classical data via $R_y(\varphi)$ single-qubit rotations on each wire is used, where the original pixel value $[0,1]$ is scaled to $\varphi \in [0, \pi]$. Then, we have a variational circuit part, which consists of 4 single-qubit rotations, parameterized with trainable weights, as well as three CNOT gates. At the end of the circuit, we measure the expectation value $\langle \sigma_z \rangle$ of the Pauli-Z operator on each qubit. Each channel is a picture with 4 $\times$ 4 pixels. After that, four output channels are flattened and fed into a fully connected layer, which yields a digit's probability.

\subsubsection{Training and results}

In this section, we describe the training process. In order to reduce the training time of the HQNN-Quanv, only $600$ images from the MNIST dataset \ref{MNIST_dataset} are used with $500$ of them acting as training data and $100$ as test data. We also use PyTorch's resize transform with bilinear interpolation to downscale images from $28 \times 28$ to $14 \times 14$ pixels. We still use a cross-entropy loss function.

While the classical model has only one way of training weights via backpropagation, the HQNN has several options, such as the parameter-shift rule, adjoint differentiation~\cite{efficient_calculation} or backpropagation (which, of course, is impossible on a real quantum computer). Adjoint differentiation seems to have the most favourable scaling with both layers and wires~\cite{Luo2020yaojlextensible}, but on this particular circuit (Fig.~\ref{fig:quanv}) backpropagation proved to be quicker.

Considering everything stated above, let us see the results of the training. We trained two CNNs with different numbers of output channels and one HQNN for 20 epochs (Fig.~\ref{fig:quanv_accuracy}(a-b)). The models were intentionally made simple and had sufficiently few parameters so as to avoid overfitting on the relatively small dataset. Test accuracies of these models are presented in Fig.~\ref{fig:quanv_accuracy}(c). For each epoch, the accuracy is averaged over 10 models with random initial weights. The error bars depict one standard deviation.

At the end of the training, the HQNN-Quanv had a test accuracy of $0.67 \pm 0.01$, which is close enough to the CNN4 result of $0.66 \pm 0.02$, while CNN1 had $0.53 \pm 0.02$. The HQNN model has only $4$ trainable weights in its quanvolutional kernel, which parameterizes rotation gates in the PQC. CNN1 and CNN4 have $4$ and $16$ trainable parameters in their convolutional kernels, respectively. Therefore, the HQNN's performance based on the accuracy score is equivalent to CNN4's, which has four times many weights in its kernel.

\section*{Discussion}\label{Discussion}

In this work, we introduced two hybrid approaches to image classification. The first approach was an HQNN-Parallel. This method allowed us to classify handwritten images of digits from the MNIST dataset with an accuracy of more than $99\%$. The classical model achieved a similar performance of $98.71\%$ and has eight times more weights in a neural network. We also tested this model on the Medical MNIST, where we achieved a quality of over 99\%, and on CIFAR-10, where we showed that the hybrid model classifies images with an accuracy of over 82\% better than its classical counterpart with a similar number of weights. These examples confirm the generalizability of the HQNN-Parallel model. Also, the successful implementation of parallel parameterized quantum circuits in the hybrid model was demonstrated, which led to such remarkable results. Our proposed architecture is a unique combination of classical and quantum layers, which we believe to be a breakthrough in solving image classification problems. 

The second approach we presented was an HQNN-Quanv. The quanvolutional layer uses fewer weights, four times less than the classical analog, to achieve approximately the same classification accuracy ($67 \pm 1\%$ for the hybrid model versus $66 \pm 2\%$ for the classical one on the test samples when averaged over ten models), while the classical analog with the same number of variational parameters as the hybrid model achieves an accuracy of $53 \pm 2\%$. 

Hybrid quantum approaches developed in this work often had significantly fewer weights in the corresponding neural networks. However, the reduced number of weights does not imply increased efficiency of the hybrid approach due to the slower training of the hybrid model compared to the classical one with the same number of weights. In practice, to obtain a practical advantage with the fewer number of weights, more efficient quantum computers or simulators of quantum computers are required.

Our research, conducted during the NISQ era, navigated the inherent constraints of current quantum circuits, such as noise levels and qubit entanglement limitations. Relying on a hybrid quantum-classical approach, the study was designed with current quantum hardware in mind and assumed seamless integration of quantum circuits within classical layers. Despite these considerations, our models demonstrated quantum superiority across three datasets, including MNIST. However, we recognize the need for broader validation to ensure holistic generalization across diverse datasets and real-world scenarios. Further research is needed to explore the full potential of HQNNs for image classification, including testing more complex architectures. Additionally, the development of more efficient optimization techniques for training PQCs and the implementation of larger-scale quantum hardware could lead to even more significant performance improvements.

In summary, our developments provide two hybrid approaches to image classification that demonstrate the power of combining classical and quantum methods. Our proposed models show improved performance over classical models with similar architectures. We believe that these results pave the way for further research in developing hybrid models that utilize the strengths of both classical and quantum computing. 

\bibliography{lib}

\begin{thebibliography}{10}

\bibitem{10.48550/arxiv.1506.01497}
Shaoqing Ren, Kaiming He, Ross Girshick, and Jian Sun.
\newblock {Faster R-CNN: Towards Real-Time Object Detection with Region
  Proposal Networks}.
\newblock {\em arXiv preprint arXiv:1506.01497}, 2015.

\bibitem{litjens2017survey}
Geert Litjens, Thijs Kooi, Babak~Ehteshami Bejnordi, Arnaud Arindra~Adiyoso
  Setio, Francesco Ciompi, Mohsen Ghafoorian, Jeroen A.W.M. van~der Laak,
  Bram~van Ginneken, and Clara~I. Sánchez.
\newblock {A survey on deep learning in medical image analysis}.
\newblock {\em Medical Image Analysis}, 42:60--88, 2017.

\bibitem{zhou2020review}
Lei Tian, Brady Hunt, Muyinatu A.~Lediju Bell, Ji~Yi, Jason~T. Smith, Marien
  Ochoa, Xavier Intes, and Nicholas~J. Durr.
\newblock {Deep Learning in Biomedical Optics}.
\newblock {\em Lasers in Surgery and Medicine}, 53(6):748--775, 2021.

\bibitem{lecun2015deep}
Yann LeCun, Yoshua Bengio, and Geoffrey Hinton.
\newblock {Deep learning}.
\newblock {\em Nature}, 521(7553):436--444, 2015.

\bibitem{segnet}
Vijay Badrinarayanan, Alex Kendall, and Roberto Cipolla.
\newblock {SegNet: A Deep Convolutional Encoder-Decoder Architecture for Image
  Segmentation}.
\newblock {\em arXiv preprint arXiv:1511.00561}, 2015.

\bibitem{densenet}
Gao Huang, Zhuang Liu, Laurens van~der Maaten, and Kilian~Q Weinberger.
\newblock {Densely Connected Convolutional Networks}.
\newblock {\em arXiv preprint arXiv:1608.06993}, 2016.

\bibitem{krizhevsky2012imagenet}
Alex Krizhevsky, Ilya Sutskever, and Geoffrey~E Hinton.
\newblock Imagenet classification with deep convolutional neural networks.
\newblock In F.~Pereira, C.J. Burges, L.~Bottou, and K.Q. Weinberger, editors,
  {\em Advances in Neural Information Processing Systems}, volume~25. Curran
  Associates, Inc., 2012.

\bibitem{redmon2017yolo9000}
Joseph Redmon and Ali Farhadi.
\newblock {YOLO9000: Better, Faster, Stronger}.
\newblock {\em arXiv preprint arXiv:1612.08242}, 2016.

\bibitem{ladd2010quantum}
Thaddeus~D. Ladd, Fedor Jelezko, Raymond Laflamme, Yasunobu Nakamura,
  Christopher~R. Monroe, and Jeremy~Lloyd O'Brien.
\newblock Quantum computers.
\newblock {\em Nature}, 464:45--53, 2010.

\bibitem{Schuld_2014}
Maria Schuld, Ilya Sinayskiy, and Francesco Petruccione.
\newblock {An introduction to quantum machine learning}.
\newblock {\em Contemporary Physics}, 56(2):172--185, 2015.

\bibitem{Biamonte2017}
Jacob Biamonte, Peter Wittek, Nicola Pancotti, Patrick Rebentrost, Nathan
  Wiebe, and Seth Lloyd.
\newblock {Quantum machine learning}.
\newblock {\em Nature}, 549(7671):195--202, 2017.

\bibitem{Dunjko.Briegel.2018}
Vedran Dunjko and Hans~J Briegel.
\newblock {Machine learning \& artificial intelligence in the quantum domain: a
  review of recent progress}.
\newblock {\em Reports on Progress in Physics}, 81(7):074001, 00 2018.

\bibitem{qml_review_2023}
Alexey Melnikov, Mohammad Kordzanganeh, Alexander Alodjants, and Ray-Kuang Lee.
\newblock Quantum machine learning: from physics to software engineering.
\newblock {\em Advances in Physics: X}, 8(1):2165452, 2023.

\bibitem{cong2019quantum}
Iris Cong, Soonwon Choi, and Mikhail~D Lukin.
\newblock Quantum convolutional neural networks.
\newblock {\em Nature Physics}, 15(12):1273--1278, 2019.

\bibitem{hur2022quantum}
Tak Hur, Leeseok Kim, and Daniel~K Park.
\newblock Quantum convolutional neural network for classical data
  classification.
\newblock {\em Quantum Machine Intelligence}, 4(1):3, 2022.

\bibitem{herrmann2022realizing}
Johannes Herrmann, Sergi~Masot Llima, Ants Remm, Petr Zapletal, Nathan~A
  McMahon, Colin Scarato, Fran{\c{c}}ois Swiadek, Christian~Kraglund Andersen,
  Christoph Hellings, Sebastian Krinner, et~al.
\newblock Realizing quantum convolutional neural networks on a superconducting
  quantum processor to recognize quantum phases.
\newblock {\em Nature Communications}, 13(1):4144, 2022.

\bibitem{Schuld2018}
Maria Schuld, Mark Fingerhuth, and Francesco Petruccione.
\newblock {Implementing a distance-based classifier with a quantum interference
  circuit}.
\newblock {\em arXiv preprint arXiv:1703.10793}, 2017.

\bibitem{fphy_2022_1069985}
Denis Bokhan, Alena~S. Mastiukova, Aleksey~S. Boev, Dmitrii~N. Trubnikov, and
  Aleksey~K. Fedorov.
\newblock {Multiclass classification using quantum convolutional neural
  networks with hybrid quantum-classical learning}.
\newblock {\em Frontiers in Physics}, 10:1069985, 2022.

\bibitem{Lloyd2013}
Seth Lloyd, Masoud Mohseni, and Patrick Rebentrost.
\newblock {Quantum algorithms for supervised and unsupervised machine
  learning}.
\newblock {\em arXiv preprint arXiv:1307.0411}, 2013.

\bibitem{365700}
P.W. Shor.
\newblock {Algorithms for quantum computation: discrete logarithms and
  factoring}.
\newblock {\em Proceedings 35th Annual Symposium on Foundations of Computer
  Science}, pages 124--134, 1994.

\bibitem{Lloyd1996UniversalQS}
Seth Lloyd.
\newblock Universal quantum simulators.
\newblock {\em Science}, 273:1073--1078, 1996.

\bibitem{Aaronson2015}
Scott Aaronson and Lijie Chen.
\newblock {Complexity-Theoretic Foundations of Quantum Supremacy Experiments}.
\newblock {\em arXiv preprint arXiv:1612.05903}, 2016.

\bibitem{Wiebe2015}
Nathan Wiebe, Ashish Kapoor, and Krysta~M Svore.
\newblock {Quantum Deep Learning}.
\newblock {\em arXiv preprint arXiv:1412.3489}, 2014.

\bibitem{Havlicek2019}
Vojtěch Havlíček, Antonio~D. Córcoles, Kristan Temme, Aram~W. Harrow,
  Abhinav Kandala, Jerry~M. Chow, and Jay~M. Gambetta.
\newblock {Supervised learning with quantum-enhanced feature spaces}.
\newblock {\em Nature}, 567(7747):209–212, 2019.

\bibitem{liu2021hybrid}
Junhua Liu, Kwan~Hui Lim, Kristin~L Wood, Wei Huang, Chu Guo, and He-Liang
  Huang.
\newblock Hybrid quantum-classical convolutional neural networks.
\newblock {\em Science China Physics, Mechanics \& Astronomy}, 64(9):290311,
  2021.

\bibitem{li2022image}
Wei Li, Peng-Cheng Chu, Guang-Zhe Liu, Yan-Bing Tian, Tian-Hui Qiu, and Shu-Mei
  Wang.
\newblock An image classification algorithm based on hybrid quantum classical
  convolutional neural network.
\newblock {\em Quantum Engineering}, 2022, 2022.

\bibitem{houssein2022hybrid}
Essam~H Houssein, Zainab Abohashima, Mohamed Elhoseny, and Waleed~M Mohamed.
\newblock {Hybrid quantum-classical convolutional neural network model for
  COVID-19 prediction using chest X-ray images}.
\newblock {\em Journal of Computational Design and Engineering}, 9(2):343--363,
  2022.

\bibitem{10.1016/j.optcom.2023.129287}
Shui-Yuan Huang, Wan-Jia An, De-Shun Zhang, and Nan-Run Zhou.
\newblock {Image classification and adversarial robustness analysis based on
  hybrid quantum–classical convolutional neural network}.
\newblock {\em Optics Communications}, 533:129287, 2023.

\bibitem{sagingalieva2023hyperparameter}
Asel Sagingalieva, Andrii Kurkin, Artem Melnikov, Daniil Kuhmistrov, et~al.
\newblock Hybrid quantum {ResNet} for car classification and its hyperparameter
  optimization.
\newblock {\em Quantum Machine Intelligence}, 5(2):38, 2023.

\bibitem{10.1063/5.0138021}
Yumin Dong, Yanying Fu, Hengrui Liu, Xuanxuan Che, Lina Sun, and Yi~Luo.
\newblock {An improved hybrid quantum-classical convolutional neural network
  for multi-class brain tumor MRI classification}.
\newblock {\em Journal of Applied Physics}, 133(6):064401, 2023.

\bibitem{10.1093/jcde/qwac003}
Essam~H Houssein, Zainab Abohashima, Mohamed Elhoseny, and Waleed~M Mohamed.
\newblock {Hybrid quantum-classical convolutional neural network model for
  COVID-19 prediction using chest X-ray images}.
\newblock {\em Journal of Computational Design and Engineering}, 9(2):343--363,
  2022.

\bibitem{zhou2023hybrid}
Nan-Run Zhou, Tian-Feng Zhang, Xin-Wen Xie, and Jun-Yun Wu.
\newblock Hybrid quantum--classical generative adversarial networks for image
  generation via learning discrete distribution.
\newblock {\em Signal Processing: Image Communication}, 110:116891, 2023.

\bibitem{tsang2022hybrid}
Shu~Lok Tsang, Maxwell~T West, Sarah~M Erfani, and Muhammad Usman.
\newblock Hybrid quantum-classical generative adversarial network for high
  resolution image generation.
\newblock {\em arXiv preprint arXiv:2212.11614}, 2022.

\bibitem{huang2021experimental}
He-Liang Huang, Yuxuan Du, Ming Gong, Youwei Zhao, Yulin Wu, Chaoyue Wang,
  Shaowei Li, Futian Liang, Jin Lin, Yu~Xu, et~al.
\newblock Experimental quantum generative adversarial networks for image
  generation.
\newblock {\em Physical Review Applied}, 16(2):024051, 2021.

\bibitem{rudolph2022generation}
Manuel~S Rudolph, Ntwali~Bashige Toussaint, Amara Katabarwa, Sonika Johri,
  Borja Peropadre, and Alejandro Perdomo-Ortiz.
\newblock Generation of high-resolution handwritten digits with an ion-trap
  quantum computer.
\newblock {\em Physical Review X}, 12(3):031010, 2022.

\bibitem{boston-housing}
Michael Perelshtein, Asel Sagingalieva, Karan Pinto, Vishal Shete, Alexey
  Pakhomchik, et~al.
\newblock {Practical application-specific advantage through hybrid quantum
  computing}.
\newblock {\em arXiv preprint arXiv:2205.04858}, 2022.

\bibitem{10.48550/arxiv.2109.02862}
Mahabubul Alam, Satwik Kundu, Rasit~Onur Topaloglu, and Swaroop Ghosh.
\newblock {Quantum-Classical Hybrid Machine Learning for Image Classification
  (ICCAD Special Session Paper)}.
\newblock {\em arXiv preprint arXiv:2109.02862}, 2021.

\bibitem{10.1109/idaacs53288.2021.9661011}
Yevhenii Trochun, Sergii Stirenko, Oleksandr Rokovyi, Oleg Alienin, Evgen
  Pavlov, and Yuri Gordienko.
\newblock {Hybrid Classic-Quantum Neural Networks for Image Classification}.
\newblock {\em 2021 11th IEEE International Conference on Intelligent Data
  Acquisition and Advanced Computing Systems: Technology and Applications
  (IDAACS)}, 2:968--972, 2021.

\bibitem{marshall2022high}
Simon~C Marshall, Casper Gyurik, and Vedran Dunjko.
\newblock High dimensional quantum learning with small quantum computers.
\newblock {\em arXiv preprint arXiv:2203.13739}, 2022.

\bibitem{kordzanganeh2023exponentially}
Mo~Kordzanganeh, Pavel Sekatski, Leonid Fedichkin, and Alexey Melnikov.
\newblock An exponentially-growing family of universal quantum circuits.
\newblock {\em Machine Learning: Science and Technology}, 4(3):035036, 2023.

\bibitem{perez2022reduce}
Adri{\'a}n P{\'e}rez-Salinas, Radoica Dra{\v{s}}ki{\'c}, Jordi Tura, and Vedran
  Dunjko.
\newblock Reduce-and-chop: Shallow circuits for deeper problems.
\newblock {\em arXiv preprint arXiv:2212.11862}, 2022.

\bibitem{kordzanganeh2023parallel}
Mo~Kordzanganeh, Daria Kosichkina, and Alexey Melnikov.
\newblock Parallel hybrid networks: an interplay between quantum and classical
  neural networks.
\newblock {\em Intelligent Computing}, 2:0028, 2023.

\bibitem{jerbi2023quantum}
Sofiene Jerbi, Lukas~J Fiderer, Hendrik Poulsen~Nautrup, Jonas~M K{\"u}bler,
  Hans~J Briegel, and Vedran Dunjko.
\newblock Quantum machine learning beyond kernel methods.
\newblock {\em Nature Communications}, 14(1):517, 2023.

\bibitem{li2020quantum}
YaoChong Li, Ri-Gui Zhou, RuQing Xu, Jia Luo, and WenWen Hu.
\newblock A quantum deep convolutional neural network for image recognition.
\newblock {\em Quantum Science and Technology}, 5(4):044003, 2020.

\bibitem{Mitarai2020}
Kosuke Mitarai, Makoto Negoro, Masahiro Kitagawa, and Keisuke Fujii.
\newblock {Quantum Circuit Learning}.
\newblock {\em Physical Review A}, 98(3):032309, 2018.

\bibitem{domingo2023hybrid}
L~Domingo, M~Djukic, C~Johnson, and F~Borondo.
\newblock Hybrid quantum-classical convolutional neural networks to improve
  molecular protein binding affinity predictions.
\newblock {\em arXiv preprint arXiv:2301.06331}, 2023.

\bibitem{pharma}
Asel Sagingalieva, Mohammad Kordzanganeh, Nurbolat Kenbayev, Daria Kosichkina,
  Tatiana Tomashuk, and Alexey Melnikov.
\newblock Hybrid quantum neural network for drug response prediction.
\newblock {\em Cancers}, 15(10):2705, 2023.

\bibitem{jain2022hybrid}
Prateek Jain and Srinjoy Ganguly.
\newblock Hybrid quantum generative adversarial networks for molecular
  simulation and drug discovery.
\newblock {\em arXiv preprint arXiv:2212.07826}, 2022.

\bibitem{sedykh2023quantum}
Alexandr Sedykh, Maninadh Podapaka, Asel Sagingalieva, Nikita Smertyak, Karan
  Pinto, Markus Pflitsch, and Alexey Melnikov.
\newblock Quantum physics-informed neural networks for simulating computational
  fluid dynamics in complex shapes.
\newblock {\em arXiv preprint arXiv:2304.11247}, 2023.

\bibitem{kurkin2023forecasting}
Andrii Kurkin, Jonas Hegemann, Mo~Kordzanganeh, and Alexey Melnikov.
\newblock Forecasting the steam mass flow in a powerplant using the parallel
  hybrid network.
\newblock {\em arXiv preprint arXiv:2307.09483}, 2023.

\bibitem{10.1016/j.eswa.2022.116583}
Eric Paquet and Farzan Soleymani.
\newblock {QuantumLeap: Hybrid quantum neural network for financial
  predictions}.
\newblock {\em Expert Systems with Applications}, 195:116583, 2022.

\bibitem{rainjonneau2023quantum}
Serge Rainjonneau, Igor Tokarev, Sergei Iudin, Saaketh Rayaprolu, Karan Pinto,
  Daria Lemtiuzhnikova, Miras Koblan, Egor Barashov, Mo~Kordzanganeh, Markus
  Pflitsch, and Alexey Melnikov.
\newblock Quantum algorithms applied to satellite mission planning for {E}arth
  observation.
\newblock {\em IEEE Journal of Selected Topics in Applied Earth Observations
  and Remote Sensing}, 16:7062--7075, 2023.

\bibitem{haboury2023supervised}
Nathan Haboury, Mo~Kordzanganeh, Sebastian Schmitt, Ayush Joshi, Igor Tokarev,
  Lukas Abdallah, Andrii Kurkin, Basil Kyriacou, and Alexey Melnikov.
\newblock A supervised hybrid quantum machine learning solution to the
  emergency escape routing problem.
\newblock {\em arXiv preprint arXiv:2307.15682}, 2023.

\bibitem{Farhi.Neven.2018mcl}
Edward Farhi and Hartmut Neven.
\newblock {Classification with Quantum Neural Networks on Near Term
  Processors}.
\newblock {\em arXiv preprint arXiv:1802.06002}, 2018.

\bibitem{Baek.Kim.2022}
Hankyul Baek, Won~Joon Yun, and Joongheon Kim.
\newblock {Scalable Quantum Convolutional Neural Networks}.
\newblock {\em arXiv preprint arXiv:2209.12372}, 2022.

\bibitem{chen2023quantum}
Guoming Chen, Qiang Chen, Shun Long, Weiheng Zhu, Zeduo Yuan, and Yilin Wu.
\newblock Quantum convolutional neural network for image classification.
\newblock {\em Pattern Analysis and Applications}, 26(2):655--667, 2023.

\bibitem{MNIST}
Li~Deng.
\newblock The {MNIST} database of handwritten digit images for machine learning
  research.
\newblock {\em IEEE Signal Processing Magazine}, 29(6):141--142, 2012.

\bibitem{Cire_an_2010}
Dan~Claudiu Cireşan, Ueli Meier, Luca~Maria Gambardella, and Jürgen
  Schmidhuber.
\newblock {Deep, Big, Simple Neural Nets for Handwritten Digit Recognition}.
\newblock {\em Neural Computation}, 22(12):3207--3220, 2010.

\bibitem{10.1109/cvpr.2012.6248110}
D~Ciresan, U~Meier, and J~Schmidhuber.
\newblock {Multi-column Deep Neural Networks for Image Classification}.
\newblock {\em 2012 IEEE Conference on Computer Vision and Pattern
  Recognition}, 1:3642--3649, 2012.

\bibitem{10.48550/arxiv.2303.05860}
Meghashrita Das and Tirupati Bolisetti.
\newblock {Variational Quantum Neural Networks (VQNNS) in Image
  Classification}.
\newblock {\em arXiv preprint arXiv:2303.05860}, 2023.

\bibitem{landman2022quantum}
Jonas Landman, Natansh Mathur, Yun~Yvonna Li, Martin Strahm, Skander Kazdaghli,
  Anupam Prakash, and Iordanis Kerenidis.
\newblock Quantum methods for neural networks and application to medical image
  classification.
\newblock {\em Quantum}, 6:881, 2022.

\bibitem{MedicalMNIST}
apolanco3225.
\newblock Medical {MNIST} classification.
\newblock \url{https://github.com/apolanco3225/Medical-MNIST-Classification},
  2017.

\bibitem{Krizhevsky09}
A.~Krizhevsky and G.~Hinton.
\newblock Learning multiple layers of features from tiny images.
\newblock {\em Master's thesis, Department of Computer Science, University of
  Toronto}, 2009.

\bibitem{Agarap2018DeepLU}
Abien~Fred Agarap.
\newblock Deep learning using rectified linear units ({ReLU}).
\newblock {\em arXiv preprint arXiv:1803.08375}, 2018.

\bibitem{https://doi.org/10.48550/arxiv.1502.03167}
Sergey Ioffe and Christian Szegedy.
\newblock Batch normalization: Accelerating deep network training by reducing
  internal covariate shift.
\newblock {\em arXiv preprint arXiv:1502.03167}, 2015.

\bibitem{tan2019mixconv}
Mingxing Tan and Quoc~V Le.
\newblock Mixconv: Mixed depthwise convolutional kernels.
\newblock {\em arXiv preprint arXiv:1907.09595}, 2019.

\bibitem{tan2019mnasnet}
Mingxing Tan, Bo~Chen, Ruoming Pang, Vijay Vasudevan, Mark Sandler, Andrew
  Howard, and Quoc~V Le.
\newblock Mnasnet: Platform-aware neural architecture search for mobile.
\newblock In {\em Proceedings of the IEEE/CVF conference on computer vision and
  pattern recognition}, pages 2820--2828, 2019.

\bibitem{lecun1998gradient}
Yann LeCun, L{\'e}on Bottou, Yoshua Bengio, and Patrick Haffner.
\newblock Gradient-based learning applied to document recognition.
\newblock {\em Proceedings of the IEEE}, 86(11):2278--2324, 1998.

\bibitem{PhysRev.70.460}
F.~Bloch.
\newblock {Nuclear Induction}.
\newblock {\em Physical Review}, 70(7-8):460--474, 1946.

\bibitem{Barenco_1995}
Adriano Barenco, Charles~H. Bennett, Richard Cleve, David~P. DiVincenzo, Norman
  Margolus, Peter Shor, Tycho Sleator, John~A. Smolin, and Harald Weinfurter.
\newblock {Elementary gates for quantum computation}.
\newblock {\em Physical Review A}, 52(5):3457--3467, 1995.

\bibitem{Rumelhart1986LearningIR}
David~E. Rumelhart, Geoffrey~E. Hinton, and Ronald~J. Williams.
\newblock {Learning representations by back-propagating errors}.
\newblock {\em Nature}, 323(6088):533--536, 10 1986.
\newblock Number: 6088 Publisher: Nature Publishing Group.

\bibitem{NEURIPS2019_9015}
Adam Paszke, Sam Gross, Francisco Massa, Adam Lerer, James Bradbury, et~al.
\newblock Pytorch: An imperative style, high-performance deep learning library.
\newblock In {\em Advances in Neural Information Processing Systems 32}, pages
  8024--8035. Curran Associates, Inc., 2019.

\bibitem{Bergholm.Killoran.2018}
Ville Bergholm, Josh Izaac, Maria Schuld, Christian Gogolin, M~Sohaib Alam,
  et~al.
\newblock {PennyLane: Automatic differentiation of hybrid quantum-classical
  computations}.
\newblock {\em arXiv preprint arXiv:1811.04968}, 2018.

\bibitem{Wierichs_2022}
David Wierichs, Josh Izaac, Cody Wang, and Cedric Yen-Yu Lin.
\newblock {General parameter-shift rules for quantum gradients}.
\newblock {\em Quantum}, 6:677, 2022.

\bibitem{benchmarkingQuantum}
Mohammad Kordzanganeh, Markus Buchberger, Basil Kyriacou, Maxim Povolotskii,
  Wilhelm Fischer, Andrii Kurkin, Wilfrid Somogyi, Asel Sagingalieva, Markus
  Pflitsch, and Alexey Melnikov.
\newblock Benchmarking simulated and physical quantum processing units using
  quantum and hybrid algorithms.
\newblock {\em Advanced Quantum Technologies}, 6:2300043, 2023.

\bibitem{Quanvolutional_networks}
Maxwell Henderson, Samriddhi Shakya, Shashindra Pradhan, and Tristan Cook.
\newblock {Quanvolutional Neural Networks: Powering Image Recognition with
  Quantum Circuits}.
\newblock {\em arXiv preprint arXiv:1904.04767}, 2019.

\bibitem{efficient_calculation}
Tyson Jones and Julien Gacon.
\newblock Efficient calculation of gradients in classical simulations of
  variational quantum algorithms.
\newblock {\em arXiv preprint arXiv:2009.02823}, 2020.

\bibitem{Luo2020yaojlextensible}
Xiu-Zhe Luo, Jin-Guo Liu, Pan Zhang, and Lei Wang.
\newblock Yao.jl: {E}xtensible, {E}fficient {F}ramework for {Q}uantum
  {A}lgorithm {D}esign.
\newblock {\em {Quantum}}, 4:341, October 2020.

\bibitem{zx-calculus}
Bob Coecke and Ross Duncan.
\newblock {Interacting quantum observables: categorical algebra and
  diagrammatics}.
\newblock {\em New Journal of Physics}, 13(4):043016, 2011.

\bibitem{schuld_fourier}
Maria Schuld, Ryan Sweke, and Johannes~Jakob Meyer.
\newblock {Effect of data encoding on the expressive power of variational
  quantum-machine-learning models}.
\newblock {\em Physical Review A}, 103(3):032430, 2021.

\bibitem{vandewetering2020zxcalculus}
John van~de Wetering.
\newblock {ZX-calculus for the working quantum computer scientist}.
\newblock {\em arXiv preprint arXiv:2012.13966}, 2020.

\bibitem{wang2023completeness}
Quanlong Wang.
\newblock {Completeness of the ZX-calculus}.
\newblock {\em arXiv preprint arXiv:2209.14894}, 2023.

\end{thebibliography}
\bibliographystyle{unsrt}

\appendix

\section{Theoretical Analysis}\label{sec:analysis:theory}

This section theoretically analyzes the quantum layers used in the HQNN-Parallel model in section~\ref{first_approach}. We focus on the methodologies of the ZX-calculus~\cite{zx-calculus} to explore circuit reducibility, and Fourier accessibility~\cite{schuld_fourier} to examine the data embedding strategy and expressivity.

\subsection{ZX-Calculus Reduction}\label{sec:analysis:zx}

\begin{figure*}[t]
    \centering
    \includegraphics[width=1\linewidth]{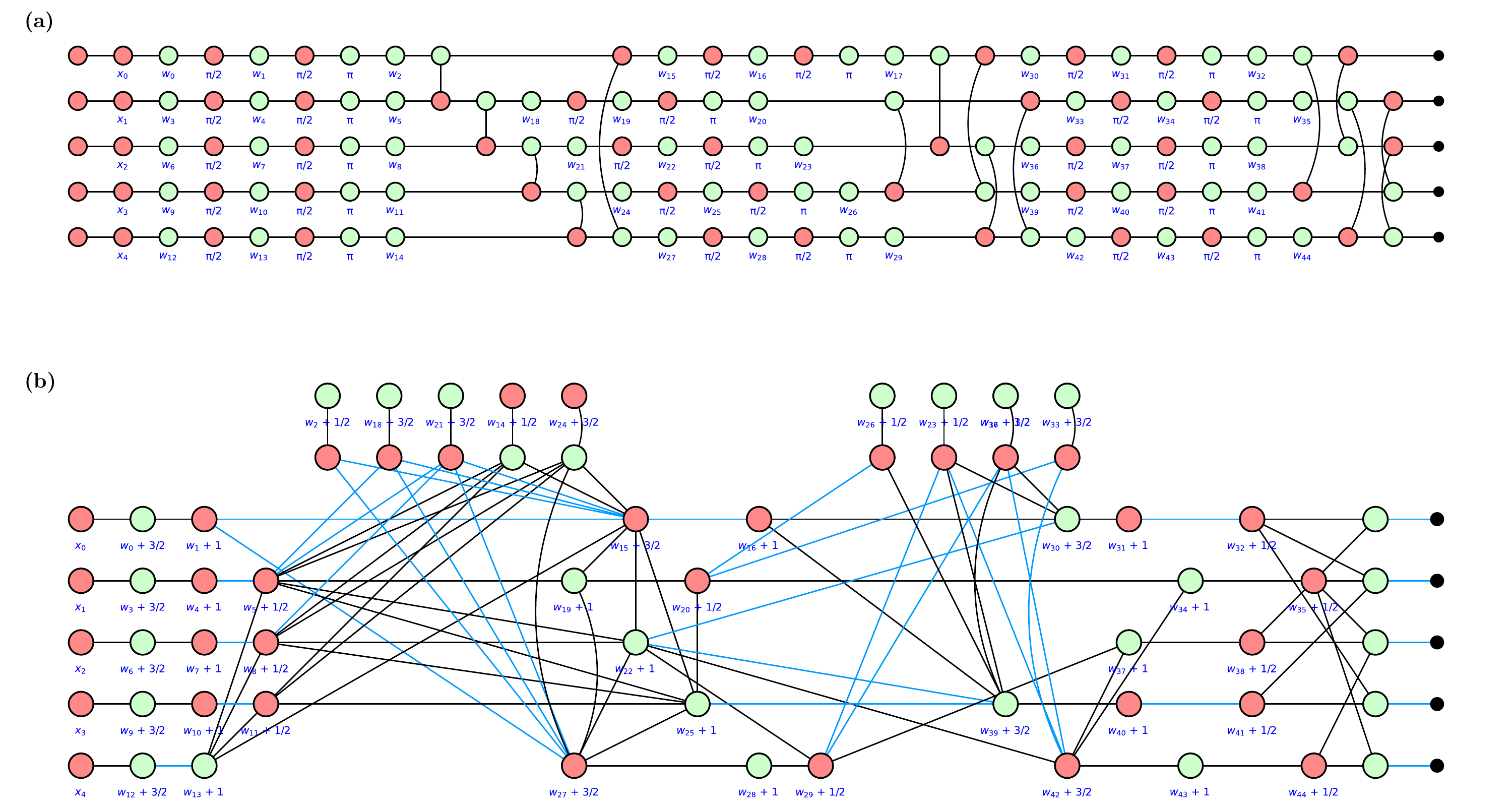}
    \caption{ZX-Calculus graphs. (a) The quantum layer from HQNN-Parallel (Fig.~\ref{fig:qfc}) written in ZX-calculus. (b) The same quantum layer as in part (a) that has been reduced with ZX rewriting rules. All weights and inputs are on independent nodes, which suggests the model is absent of redundancy.}
    \label{fig:ZX-diagram}
\end{figure*}

ZX-calculus is a graphical language that replaces circuit diagrams with ZX-diagrams by replacing quantum tensors with so-called ``spiders'', nodes on a graph with edges that connect them ~\cite{zx-calculus, vandewetering2020zxcalculus, wang2023completeness}. These spiders come in two flavors, a light or green-colored spider that represents tensors in the $Z$ basis ($\ket{0}$, $\ket{1}$) and a dark or red-colored spider that represents tensors in the $X$ basis ($\ket{+}$, $\ket{-}$). ZX-diagrams can be simplified and reduced with the language's graphical rewrite rules based on the underlying quantum operations. For example, repetitions of Pauli rotations sum together to form one Pauli rotation with an angle equal to the sum of its parts. This is translated into ZX-calculus as a specific instance of the more general rule of ``fusing'' spiders, where nodes of the same color combine and sum their angles. More generally, quantum operations often possess subtle symmetries that make it difficult to implement effective circuits, and for exponentially large systems, matrix multiplication quickly becomes unwieldy. Essentially, ZX calculus replaces tedious matrix multiplication of quantum gates with easy-to-apply graphical rules. Thus, analysis of ZX diagrams are helpful for identifying redundancies in a quantum model.

To analyze the reduced quantum layer of Fig.~\ref{fig:ZX-diagram}(a), we first represented it as a red-green ZX-diagram. Then ZX-calculus's rewriting rules are applied to simplify the circuit and remove redundancies. Finally, the resulting new circuit is extracted. Fig.~\ref{fig:ZX-diagram}(b) shows the simplified circuit in ZX form. Note that the reduced ZX-form still has all the initial parameters $w_i$ on separate nodes. This indicates that the model does not reduce away any of the inputs or trainable parameters. This verifies that all the weights and input data in the circuit actually make an impact on the final result.

\subsection{Fourier Expressivity} \label{sec:analysis:fourier}

\begin{figure*}[t]            
\centering    
     \includegraphics[width=1.0\linewidth]{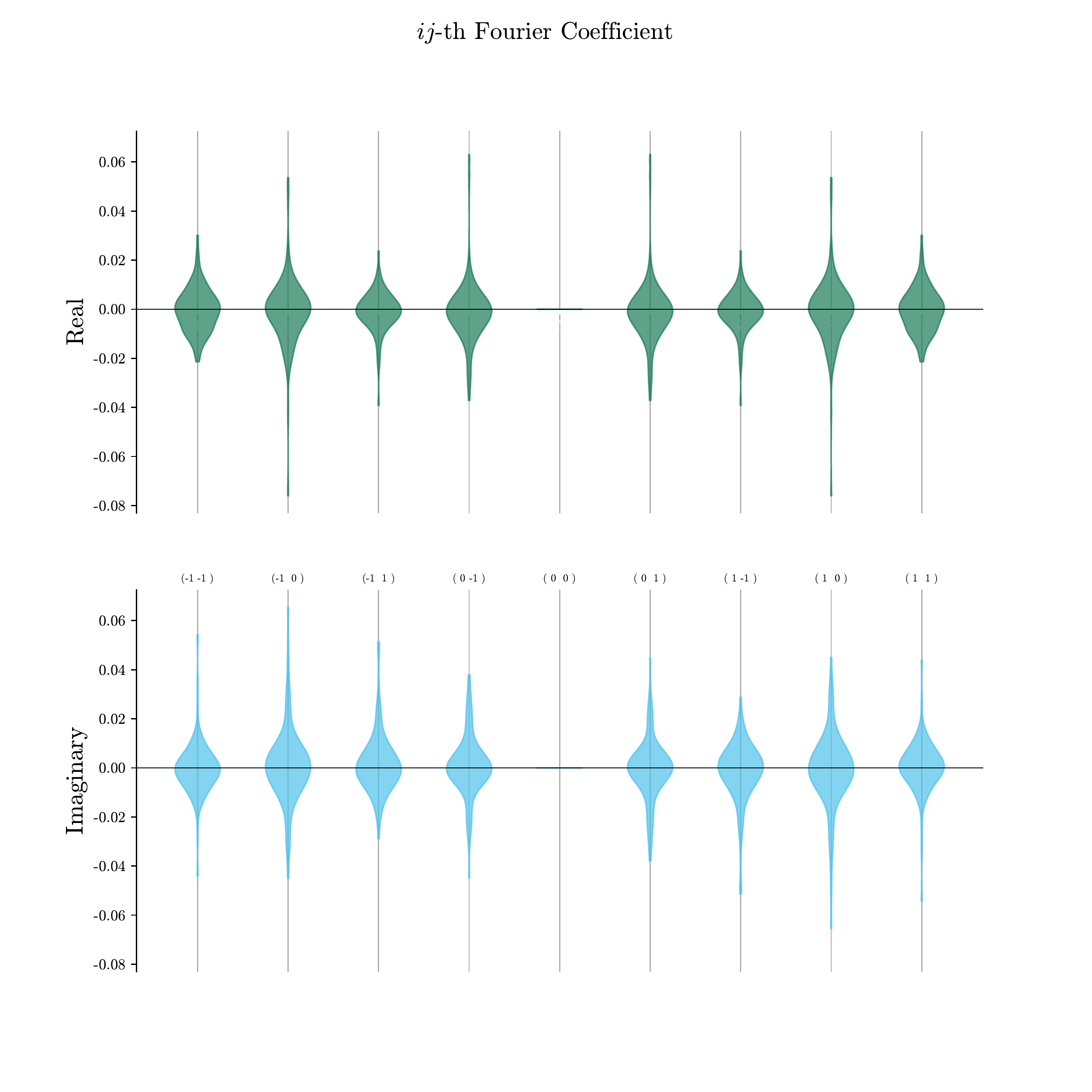}
    \caption{
     A violin chart of 100 samples of the values of the Fourier coefficients for the first and second input parameters of the final output measurement. The $ij$-th indices along the center line represent the Fourier coefficient $c_{ij}$. The width of the violins represents the number of samples at that magnitude. The large spread on both the real and imaginary part of every coefficient implies high expressivity in the model.}
    \label{fig:fourier_violin}
\end{figure*}

Ref.~\cite{schuld_fourier} showed that the output of a parameterized quantum circuit is equivalent to a truncated Fourier series. For a feature vector of length $N$, the Fourier series as a function of the feature vector $\mathbf{x}$ and trainable parameters $\mathbf{\theta}$ is:
\begin{align*}
f_\mathbf{\theta}(\mathbf{x}) = \sum_{\omega_1 \in \Omega_1} \ldots \sum_{\omega_N \in \Omega_N} c_{\omega_1 \ldots \boldsymbol{\omega}_N} (\mathbf{\theta}) e^{-i \mathbf{\omega} \cdot \mathbf{x}},
\end{align*}
where $\omega_i \in \{ -d_i, \ldots, 0 , \dots, d_i\}$. In other words, the number of terms in the Fourier series is one more than twice the number of times that input was placed in the circuit, $d$. In this analysis, we show the expressivity of the function $f_\mathbf{\theta}(\mathbf{x})$ by sampling over a uniform distribution of random values for each $\theta_i$ from $[0, 2 \pi ]$ and by sampling equidistant $x$ values with a sampling frequency of $d$.

For visual clarity, we display only the first two terms (associated with the first two features) of the model on the final output of the quantum circuit. If we write these inputs as $x$ and $y$, the output function $f_\mathbf{\theta}(\mathbf{x}) $ becomes,
\begin{align*}
f_\mathbf{\theta}(x,  y) = \sum^1_{\omega_x =-1} \sum^1_{\omega_y =-1} c_{\omega_x,\omega_y} (\mathbf{\theta}) \: e^{-i \omega_x x  } e^{-i \omega_y y}
\end{align*}

Fig.~\ref{fig:fourier_violin} demonstrates a violin plot of the Fourier coefficients $c_{\omega_x, \omega_y}$ sampled over various $\theta$ realizations. A completely non-expressive model would have terms close to zero for all these coefficients. Instead, the figure shows that the weights of the quantum layer have a wide range of possible solutions. Additionally, the determinate of the correlation matrix of all values was equal to zero for every output, demonstrating the independent and expressive nature of all the Fourier terms in the model.

\end{document}